\documentclass{article}
\usepackage{amsmath,amsfonts,graphicx,mlspconf,multirow,makecell}
\usepackage[acronym]{glossaries}
\usepackage[numbers,sort&compress]{natbib}

\copyrightnotice{978-1-7281-6338-3/21/\$31.00 {\copyright}2021 IEEE}

\toappear{2021 IEEE International Workshop on Machine Learning for Signal Processing, Oct.\ 25--28, 2021, Gold Coast, Australia}

\title{Disentangled Speech Representation Learning Based on Factorized Hierarchical Variational Autoencoder with Self-Supervised Objective} 

\name{Yuying Xie\thanks{The work of Yuying Xie is supported by China Scholarship Council.}, Thomas Arildsen, Zheng-Hua Tan}
\address{Department of Electronic Systems, Aalborg University, Aalborg, Denmark}

\begin{document}

\maketitle
\newacronym{vae}{VAE}{variational autoencoder}
\newacronym{lstm}{LSTM}{long short-term memory}
\newacronym{apc}{APC}{autoregressive predictive coding}
\newacronym{fhvae}{FHVAE}{factorized hierarchical variational autoencoder}
\newacronym{cnn}{CNN}{convolutional neural network}

\begin{abstract}
Disentangled representation learning aims to extract explanatory features or factors and retain salient information. \Acrfull{fhvae} presents a way to disentangle a speech signal into sequential-level and segmental-level features, which represent speaker identity and speech content information, respectively. As a self-supervised objective, \acrfull{apc}, on the other hand, has been used in extracting meaningful and transferable speech features for multiple downstream tasks. Inspired by the success of these two representation learning methods, this paper proposes to integrate the APC objective into the FHVAE framework aiming at benefiting from the additional self-supervision target. The main proposed method requires neither more training data nor more computational cost at test time, but obtains improved meaningful representations while maintaining disentanglement. The experiments were conducted on the TIMIT dataset. Results demonstrate that FHVAE equipped with the additional self-supervised objective is able to learn features providing superior performance for tasks including speech recognition and speaker recognition. Furthermore, voice conversion, as one application of disentangled representation learning, has been applied and evaluated. The results show performance similar to baseline of the new framework on voice conversion.
\end{abstract}
\section{Introduction}
\label{sec:intro}

Since data representation impacts subsequent algorithm performance highly~\cite{bengio2013representation}, representation learning, as the technique that learns how to extract powerful features for further data processing, becomes a very important research area. The application of representation learning is extensive, involving speech~\cite{representation-speech-recognition}, image~\cite{representation-learning-graph}, natural language precessing~\cite{representation-learning-nlp}. 
Since there are often multiple independent and explainable factors in observed data, disentangled representation learning provides a way to extract representations from observed data to symbolize these explainable factors separately.

\Acrfull{vae} \cite{KingmaD.P2014AVB} is one of the most popular generative models in the representation learning field. As an extended version of \acrshort{vae}, \acrfull{fhvae} \cite{10.5555/3294771.3294950} provides an unsupervised way to disentangle sequence data into sequential features and segmental features. Experiments in~\cite{10.5555/3294771.3294950} evaluate sequential features and segmental features, and apply FHVAE for voice conversion and robust speech recognition. A \acrfull{cnn} version of FHVAE is proposed in~\cite{Gburrek_Glarner_Ebbers_Haeb-Umbach_Wagner_2019}, and the experiments show that using \acrshort{cnn} layers in FHVAE aids latent segmental feature extraction. The idea of FHVAE has also been used in emotional voice conversion \cite{FHVAE-emotion} and singing voice conversion \cite{FHVAE_singing}. 

\Acrfull{apc}, as one neural version of predictive coding, has been proposed in~\cite{chung2019unsupervised}.
As a self-supervised objective, APC provides another speech representation learning method. The architecture of APC is autoencoder-like, but the decoder aims to predict features of future frames rather than reconstruct the current one. The strategy of the encoder in APC is to retain salient information in features extracted from the surface feature (e.g., log-Mel spectrogram) for downstream tasks. APC has shown impressive results in speech recognition, speech translation and speaker recognition~\cite{9054438}, and has competitive performance in state-of-the-art generative models according to~\cite{superb}. One successful application of APC is as a pre-training means to learn to extract transferable speech representations \cite{9054438, 9339931}. \cite{reinhold} proposes another work by using contrastive predictive coding to support factorized disentangled representation learning for speech signal.

Motivated by the impressive feature extraction ability of APC, the idea of using APC to aid FHVAE performance comes to the surface. After several attempts on framework architecture, this paper proposes a new method by using APC as an additional loss in the FHVAE framework. This method considers the balance between reconstruction and prediction, and improves feature extraction ability without requiring more training data, 
while maintaining disentanglement. The structure of encoders and decoders of the proposed model are also explored in this paper. The proposed method is evaluated on multiple dimensions. One is feature extraction ability, evaluated by speech recognition, speaker verification and speaker identification. And another is how well further tasks work on the proposed method. Voice conversion has been chosen in this paper as the application. 

The organization of this paper is as follows: Section~\ref{sec:background} briefly reviews the theory of FHVAE and APC; Section~\ref{sec:Proposed method} presents the details of the proposed method; experiments are put in Section~\ref{sec:experiment}, and conclusion is summarized in Section~\ref{sec:conclusion}.

\section{Theoretical background}
\label{sec:background}

\subsection{Factorized hierarchical variational autoencoder}

\begin{figure}[htb]
  \centering
  \centerline{\includegraphics[width=7.0cm]{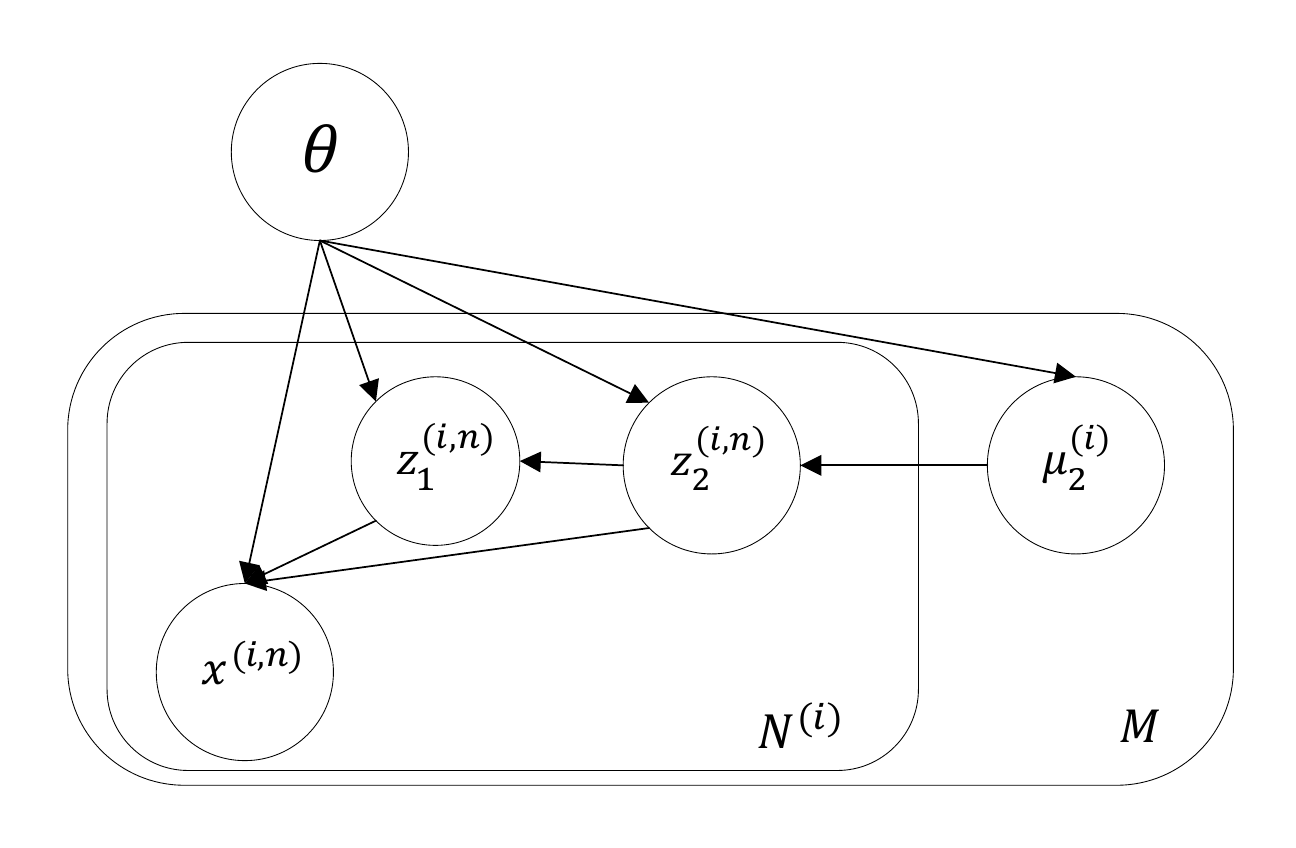}}
  \centerline{(a) generative model of FHVAE \label{fig1:a}}\medskip
\hfill
  \label{fig1:b}
  \centering
  \centerline{\includegraphics[width=7.0cm]{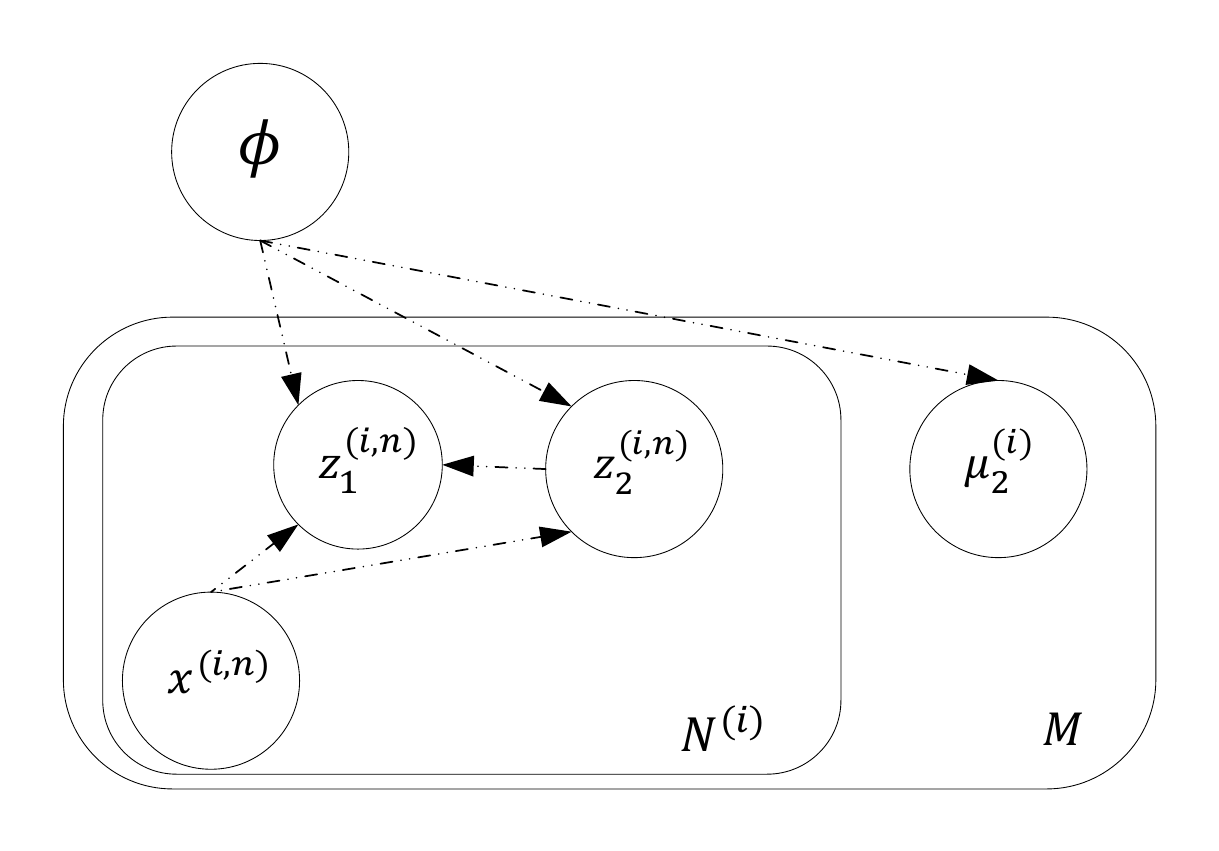}}
  \centerline{(b) inference model of FHVAE}\medskip
%
\caption{Graphical illustration of FHVAE (adpated from \cite{10.5555/3294771.3294950})}
\label{fig1}
\end{figure}

\Acrfull{fhvae} \cite{10.5555/3294771.3294950} provides a disentangled and interpretable way to represent sequence signals.
Speaker identity and linguistic content from utterances, as prominent information for later speech processing, have been considered as disentangled representations in FHVAE framework.
One sequence is composed of several segments, and several frames constitute one segment. If we use variables to represent speaker identity information, the variables should change significantly between sequences rather than within one sequence, therefore it is denoted a latent sequential variable. Since linguistic content varies in time, the relevant variables should change between segments, and is thus considered a latent segmental variable in the following. By presuming observed data $x$ is generated from some random process involving a latent sequential variable and a latent segmental variable, FHVAE unsupervisedly extracts segmental latent variable $z_1$ to represent speech content information, and sequential latent variable $z_2$ to represent speaker identity information separately. The graphical illustration is shown in Figure~\ref{fig1}. $\theta$ in Figure~\ref{fig1}(a) represents the set of parameters in the generative model, and $\phi$ in Figure~\ref{fig1}(b) denotes the parameter set in the inference model. 

Suppose a speech feature (e.g., log-Mel spectrogram) dataset $D=\{X^{(i)}\}^{M}_{i=1}$ consists of $M$ i.i.d sequences, and the $i$-th sequence $X^{(i)}$ is composed of $N^{(i)}$ segments, i.e. $X^{(i)}=\{x^{(i,n)}\}_{n=1}^{N^{(i)}}$. The \acrshort{fhvae} objective function defined on segment $x^{(i,n)}$ is as follows:
\begin{equation}
\begin{aligned}
    &\mathcal{L}_{i,n} \\&=  \mathbb{E}_{q_{\phi}(z_{1}^{(i,n)},z_{2}^{(i,n)}|x^{(i,n)})}[\log p_{\theta}(x^{(i,n)}|z_1^{(i,n)},z_2^{(i,n)})] \\
    & - \mathbb{E}_{q_{\phi}(z_2^{(i,n)}|x^{(i,n)})}[D_{KL}(q_{\phi}(z_1^{(i,n)}|x^{(i,n)},z_2^{(i,n)})\|p_{\theta}(z_1^{(i,n)}))]  \\
    & - D_{KL}(q_{\phi}(z_2^{(i,n)}|x^{(i,n)})\|p_{\theta}(z_2^{(i,n)}| \Tilde{\mu}_2^{(i)})) \\
    &  + \frac{1}{N^{(i)}}\log p _{\theta}(\Tilde{\mu}_2^{(i)})
\end{aligned}
\label{equa1}
\end{equation}

This objective function contains a reconstruction loss and a regularisation loss. $p$ and $q$ represent prior and posterior probability separately. The first term in Eq.~(\ref{equa1}) uses log-likelihood loss to measure the reconstruction loss. The second and third terms utilize Kullback–Leibler (KL) divergence to measure the distance between prior and posterior of latent variables $z_1$ and $z_2$, respectively. The fourth term in Eq.~(\ref{equa1}) is the log prior probability of $\Tilde{\mu}_2$.

As Figure~\ref{fig1}(a) shows, the generation process begins from $z_2^{(i,n)}$. The prior $p_{\theta}(z_2^{(i,n)})$ is a multivariate Gaussian distribution with fixed variance $\sigma_2$ and trainable mean $\mu_2^{(i)}$. $\mu_2^{(i)}$ is also called s-vector~\cite{10.5555/3294771.3294950}. The prior $p_{\theta}(\mu_2^{(i)})$ is $N(\mathbf{0},\mathbf{I})$, and the posterior mean of $\mu_2$ denoted as $\Tilde{\mu}_2$ is from the neural network. The prior of segmental variable $z_1$ and data $x$ are the same, i.e. $N(\mathbf{0},\mathbf{I})$. The generation and inference of segmental variable $z_1$ is dependent on sequential variable $z_2$, and thus the posterior of $z_1$ is dependent on $z_2$. The posteriors of $z_1$ and $x$ are all generated from the neural network.


\subsection{Autoregressive predictive coding}
\Acrfull{apc}, as a self-supervised objective, is another solution to learn representations for sequential data. The architecture of APC is autoencoder-like. This encoder tries to extract speech representations without a specific target and aims to retain as much information as possible. The biggest difference between APC architecture and autoencoder is, the decoder in APC does not reconstruct the input but predicts future frames of input speech feature. Speaker recognition and speech recognition experiments from~\cite{9054438} show that features extracted by APC outperform surface feature (log-Mel spectrogram) in representing speaker identity and linguistic content. Experiments in~\cite{chung2019unsupervised} also explore how features extracted from different layers perform in different tasks. Specifically, features extracted from each of 3 RNN layers have been used for speech recognition and speaker recognition. The result shows that features from shallow layers contain more speaker identity information, while from deeper layers show better performance on speech recognition.

Assume an utterance is represented by a sequence of acoustic feature vectors $X=\{x^{(n)}\}_{n=1}^{N}$ (like spectrogram or FBank). APC processes past and current frame features and outputs predictions of future frames. Paper~\cite{9054438} discusses the impact when predictive output is $m$ frames ahead of the current frame. Experiments indicate that with a too small $m$, it is hard for APC to infer more global structures, while a very large $m$ makes prediction much more challenging. The optimal $m$ for RNN-based APC framework is 3.
The objective function, using L1 loss, is shown in Eq.~\eqref{equa2}. $x$ and $y$ denote the data and predictive output, respectively.



\begin{equation}
\begin{aligned}
    \mathcal{L}_{APC} = \sum_{n=1}^{N-m}|y^{(n)}-x^{(n+m)}|
\end{aligned}
\label{equa2}
\end{equation}

\section{Proposed method}
\label{sec:Proposed method}

FHVAE presents a successful way to disentangle sequential features and segmental features in one framework. Inspired by APC's powerful ability to extract features, a new architecture is proposed here by using APC as an additional self-supervised loss in FHVAE framework. In the remainder of the paper, FHVAE-APC is used to denote the proposed method.

\begin{figure}[t]
  \centering
  \includegraphics[width=0.7\columnwidth]{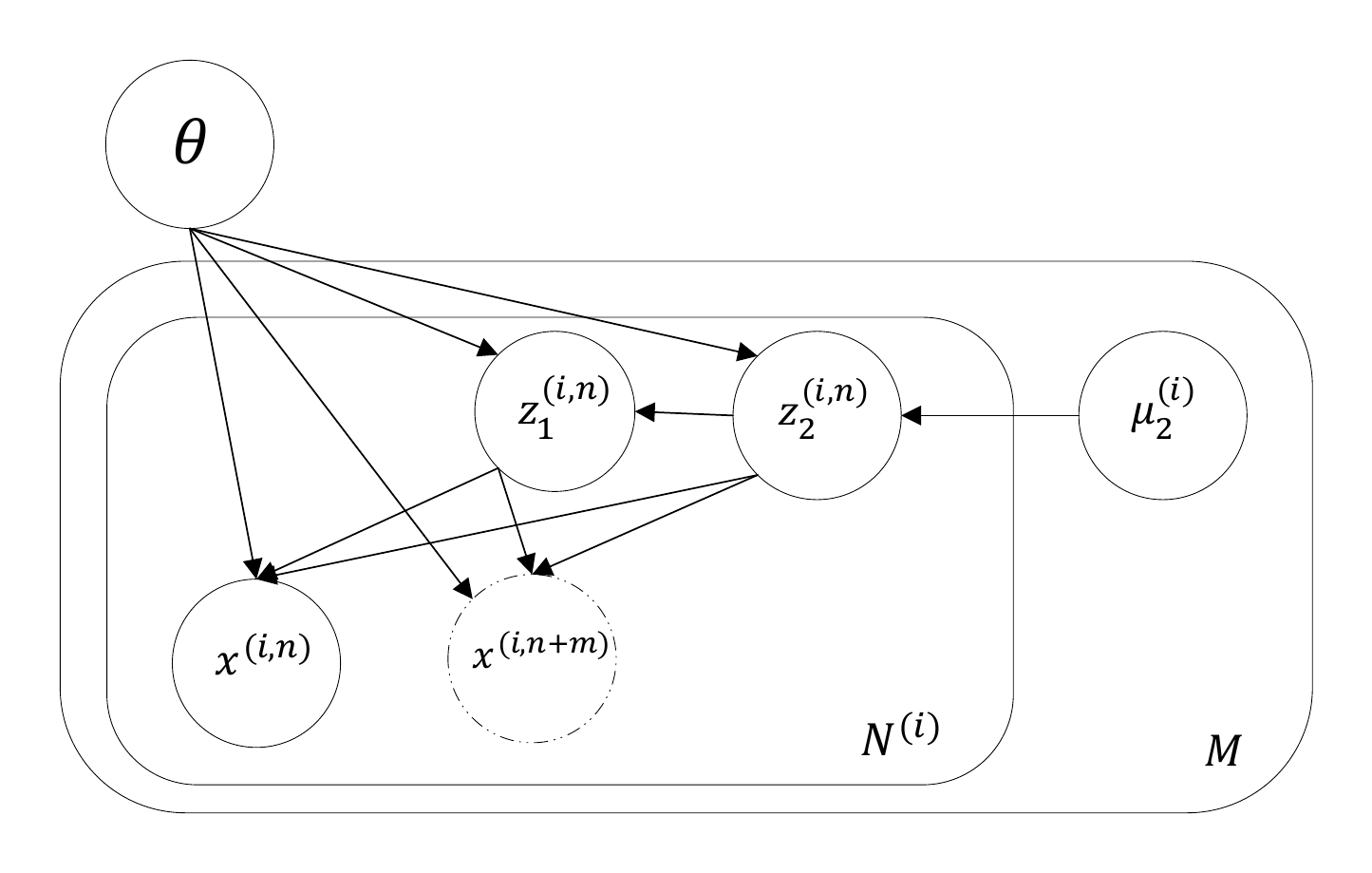}
  \caption{Generative model of FHVAE-APC. The new variable introduced in FHVAE-APC is highlighted using the dashed circle.}
  \label{fig:generative-model}
\end{figure}

The inference model in FHVAE-APC is identical to the original FHVAE~\cite{10.5555/3294771.3294950}, but not the generative model. Figure~\ref{fig:generative-model} shows the new generative model, and the difference between FHVAE and FHVAE-APC is highlighted by the dashed circle. Compared to FHVAE, the new model not only reconstructs current-frame data but also predicts future frame data simultaneously according to latent variables $z_1$ and $z_2$. It is worth noting that these two latent variables are inferred by the inference model. Since inference model and generative model have mutual effect and are trained collectively, the power of the inference model is hypothesized to improve by adding the prediction part in the generative model.

\begin{figure}[t]
  \centering
  \includegraphics[width=\columnwidth]{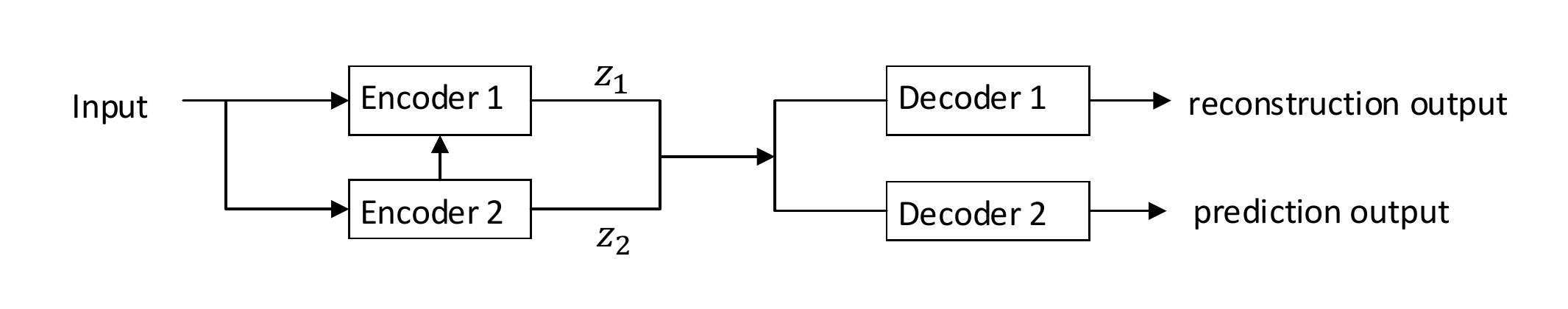}
  \caption{Architecture of FHVAE-APC.}
  \label{fig:architecture}
\end{figure}

Several attempts on improving FHVAE performance with APC loss have been made. The final proposed architecture of FHVAE-APC is similar to the original FHVAE architecture~\cite{10.5555/3294771.3294950}, but adds a parallel decoder as Figure~\ref{fig:architecture} shows. This added decoder is used for prediction, and has identical input to the original reconstruction decoder. The predictive decoder works directly on latent variables extraction in this way. 

Equation~\eqref{eq FHVAE-APC} shows the objective function of FHVAE-APC. In eq.~\eqref{eq FHVAE-APC}, $\Tilde{x}$, $x$, and $y$ denote reconstructed data from decoder 1, observed data, and predictive data from decoder 2, respectively. For balancing the two decoders in the whole framework, both reconstruction loss and prediction loss use L2 loss to measure, i.e. the first two terms in eq.~\eqref{eq FHVAE-APC}, RHS. The remaining terms in equation~\eqref{eq FHVAE-APC} are the same as in FHVAE. According to the results from~\cite{9054438} and primary experiments, $m$ equals 3 in the following experiments.

\begin{equation}
\begin{aligned}
    &\mathcal{L}_{i,n,m} \\&= 
    \| \Tilde{x}^{(i,n)} - x^{(i,n)}\|^2 + \| y^{(i,n)}-x^{(i,n+m)} \|^2 \\
    & + \mathbb{E}_{q_{\phi}(z_2^{(i,n)}|x^{(i,n)})}[D_{KL}(q_{\phi}(z_1^{(i,n)}|x^{(i,n)},z_2^{(i,n)})\|p_{\theta}(z_1^{(i,n)}))]  \\
    & + D_{KL}(q_{\phi}(z_2^{(i,n)}|x^{(i,n)})\|p_{\theta}(z_2^{(i,n)}| \Tilde{\mu}_2^{(i)})) \\
    &  - \frac{1}{N^{(i)}}\log p _{\theta}(\Tilde{\mu}_2^{(i)})
\end{aligned}
\label{eq FHVAE-APC}
\end{equation}

\section{Experiments}
\label{sec:experiment}
\begin{table*}[th]
  \caption{Segmental-level feature evaluation}
  \label{tab:z1 speech-rec}
  \centering
  \begin{tabular}{c c c c c c c}
    \hline
    \multicolumn{2}{c}{\textbf{Framework}} & 
    \textbf{WER(\%)} &  \textbf{Correct(\%)}&  \textbf{Substitution(\%)}&  \textbf{Deletion(\%)}&  \textbf{Insertion(\%)}\\
    \hline
    \multicolumn{2}{c}{FHVAE} & $28.0$ & $75.6$ & $18.4$ & $6.0$ & $3.6$   \\ \hline 
    \multirow{3}{*}{FHVAE-APC} &
    enc1-dec1 & $23.8$ & $79.3$ & $15.5$ & $5.2$ & $3.1$ ~~~ \\
    & enc2-dec1 & $26.5$ & $77.2$ & $17.4$ & $5.4$ & $3.7$ ~~~ \\
    & enc2-dec2 & $25.1$ & $78.6$ & $16.1$ & $5.3$ & $3.6$ ~~~ \\
    \hline
  \end{tabular}
\end{table*}

\subsection{Dataset}

The TIMIT dataset~\cite{garofolo1993darpa} is used in the following experiments. Particularly, TIMIT training set, development set, and core test set, which has been used, contains 462 speakers, 50 speakers, and 24 speakers respectively with no overlap in speakers among them. TIMIT provides three kinds of utterances per speaker: 2 dialect sentences (SA utterances), 5 phonetically-compact sentences (SX utterances) and 3 phonetically-diverse sentences (SI utterances). In the following experiments, only SX and SI utterances are used, i.e. 8 utterances per speaker. 200-dimensional log-magnitude spectrogram is utilized as input speech feature.

\subsection{Implementation details}
In original FHVAE~\cite{10.5555/3294771.3294950}, encoders and decoders all have similar structures which contain a LSTM layer with 256 units and a fully-connected layer of different sizes. The dimension of latent variables $z_1$ and $z_2$ equals 32. This structure has also been used as our starting point: all encoders and decoders have identical structure to FHVAE, which means the prediction decoder has the same structure as the reconstruction decoder. For short, this framework will be denoted 'FHVAE-APC enc1-dec1' in the following experiments.

Besides, \cite{chung2019unsupervised} inspires us to improve our proposed method's feature extraction ability by adding encoder layers. Experiments in~\cite{chung2019unsupervised} show that features extracted from shallow layers contain more information about speaker identity but less speech content, while features extracted from deep layers show better performance in a phone classification experiment but not a speaker verification experiment. Motivated by this, we use two LSTM layers which both have 256 hidden units and a fully-connected layer in encoder 1, but retain the structures of the $z_2$ encoder and the decoders. This model will be denoted 'FHVAE-APC enc2-dec1' in the following.
In addition, we also tried to emphasize the prediction decoder in the 'FHVAE-APC enc2-dec1' framework by using two LSTM layers with 256 units rather than one LSTM layer. This model is referred to as 'FHVAE-APC enc2-dec2'.
The training process and other parameter settings all follow~\cite{10.5555/3294771.3294950}. As decoder 2 is discarded at test time, the main proposed model ’FHVAE-APC enc1-dec1’ has the same run-time computation cost as 'FHVAE', while the other two models have larger encoders and thus increased computational cost.

\subsection{Segmental-level feature evaluation}

The feature we used in the following experiments is mean and log variance of latent segmental feature $z_1$. 
The speech recognition system implemented by the Kaldi toolbox \cite{povey2011kaldi} has been used to evaluate this segmental feature. 
The system is trained by extracted features from the TIMIT training set, and Table~\ref{tab:z1 speech-rec} summarizes the evaluation results based on features from the TIMIT core test set.

All the three FHVAE-APC model results in Table~\ref{tab:z1 speech-rec} are better than the baseline in terms of word error rate (WER), which illustrates that adding APC loss in FHVAE framework can uprate segmental feature extraction. In these three FHVAE-APC model versions, 'FHVAE-APC enc1-dec1' has the best performance, with corresponding WER of $23.8\%$. Performances of 'FHVAE-APC enc2-dec2' and 'FHVAE-APC enc2-dec1' are worse than 'FHVAE-APC enc1-dec1', even though these two models contain more layers in the $z_1$ encoder. 
The reason may be that, as the structures of the $z_1$ and $z_2$ encoder are different in the latter two frameworks, the balance of the whole framework is broken, and could cause segmental feature extraction performance reduction.

\subsection{Sequential-level feature evaluation}

The mean of $z_2$, i.e. $\mu_2$ is used as sequential-level feature in experiments. To get a more objective evaluation result, both speaker verification and speaker identification experiments have been conducted and three approaches have been applied. The speaker verification experiment calculates cosine similarity score between all sequential features from TIMIT core test set and uses equal error rate (EER) as evaluation metric. Two neural network approaches are applied in speaker identification experiment: 1-layer GRU classifier, as one of them, using Adam optimizer with learning rate equal to 0.001. Model training stops if loss does not change after 15 epochs. Besides, the other classifier for evaluation consists of 1-layer GRU with 512 units and a fully-connected (FC) layer; optimizer and learning rate are identical to the former. For less calculation cost, model training stops if loss does not decrease after 5 epochs here. Both neural network methods use cross-entropy loss. Features from the TIMIT core test set are used in these neural network experiments. Since 24 speakers are contained and each speaker has 8 utterances in core test set, 8-fold cross-validation is used here. Every time 24 features from different speakers are used as test set, another 24 features from different speakers are used as development set and the residual 144 
features are used as training set.

\begin{table}[th]
  \caption{Sequential-level feature evaluation}
  \label{tab:z2 speaker-class}
  \centering
  \begin{tabular}{ c c c c c}
    \hline
    \multicolumn{2}{c}{\multirow{2}{*}{\textbf{Framework}}} & \textbf{EER(\%)} &  \multicolumn{2}{c}{\textbf{Accuracy(\%)}}  \\
    \cline{4-5}
    && \scriptsize \makecell{cosine\\similarity}& GRU & GRU+FC \\ 
    \hline
    \multicolumn{2}{c}{FHVAE} & $4.12$ & $73.96$ & $90.10$ ~~~ \\ \hline
    \multirow{3}{*}{\makecell{FHVAE\\-APC}}
    & enc1-dec1 & $4.03$ & $78.65$ & $97.40$ ~~~ \\
    & enc2-dec1 & $2.60$ & $86.98$ & $93.75$ ~~~ \\
    & enc2-dec2 & $2.60$ & $89.58$ & $94.27$ ~~~ \\
    \hline
  \end{tabular}
\end{table}

Table \ref{tab:z2 speaker-class} shows the experiment results. FHVAE-APC models are all superior to baseline in the three experiments. Compared to baseline, 'FHVAE-APC enc1-dec1' shows slightly improved performance in EER and 1-layer GRU experiment, but obvious advantage when using 1-layer GRU and a FC layer as classifier. With an added LSTM layer in the $z_1$ encoder, FHVAE-APC shows stronger ability to extract latent segmental features in EER and 1-layer GRU experiment. The result of 'FHVAE-APC enc2-dec2' from 'GRU+FC' is a little worse than 'FHVAE-APC enc1-dec1' but still better than the baseline.

These results show that adding the APC module to the FHVAE framework improves sequential feature extraction ability. The conclusion from~\cite{chung2019unsupervised} about features from different layers placing emphasis on different speech factors has also been confirmed in these 
experiments.

\begin{table*}[th]
  \caption{Speech recognition results of voice conversion}
  \label{tab:vc speech-rec}
  \centering
  \begin{tabular}{c c c c c c c}
    \hline
    \multicolumn{2}{c}{\textbf{Framework}} & 
    \textbf{WER(\%)} &  \textbf{Correct(\%)}&  \textbf{Substitution(\%)}&  \textbf{Deletion(\%)}&  \textbf{Insertion(\%)}\\
    \hline
    \multicolumn{2}{c}{FHVAE} & $33.3$ & $70.3$ & $22.3$ & $7.4$ & $3.6$   \\ \hline 
    \multirow{3}{*}{FHVAE-APC} &
    enc1-dec1 & $33.8$ & $69.5$ & $22.0$ & $8.5$ & $3.3$ ~~~ \\
    & enc2-dec1 & $33.9$ & $69.4$ & $22.9$ & $7.7$ & $3.3$ ~~~ \\
    & enc2-dec2 & $32.5$ & $70.6$ & $21.5$ & $7.8$ & $3.1$ ~~~ \\
    \hline
  \end{tabular}
\end{table*}

\begin{table*}[th]
  \caption{Speaker verification results of voice conversion}
  \label{tab:vc speaker-recognition}
  \centering
  \begin{tabular}{ c|c|c|c|c|c|c|c|c|c|c}
    \hline
    \multicolumn{2}{c|}{\multirow{2}{*}{\textbf{Framework}}} &
    \multicolumn{3}{@{}c@{}|}{\textbf{$EER_A(\%)$}} &
    \multicolumn{3}{@{}c@{}|}{\textbf{$EER_B(\%)$}} &
    \multicolumn{3}{@{}c@{}}{\textbf{$ND$}} \\
    \cline{3-11}
    \multicolumn{2}{c|}{} &  Female & Male & All 
    &  Female & Male & All
    &  Female & Male & All\\ 
    \hline
    \multicolumn{2}{@{}c@{}|}{TIMIT core test set} & $8.33$ & $4.17$ & $4.17$ & - &  - &  - &  - &  - &  - ~~~ \\ \hline
    \multicolumn{2}{@{}c@{}|}{FHVAE} & $29.69$ & $24.61$ & $32.47$ & $32.81$ & $39.45$ & $28.99$ & $0.37$ & $3.56$ & $-0.84$  ~~~ \\ \hline
    \multirow{3}{*}{FHVAE-APC} 
    & enc1-dec1 & $25.00$ & $22.66$ & $26.74$ & $42.19$ & $39.45$ & $34.20$ & $2.06$ & $4.03$ & $1.79$ ~~~ \\
    & enc2-dec1 & $32.81$ & $24.61$ & $31.94$ & $31.25$ & $37.89$ & $29.34$ & $-0.19$ & $3.19$ & $-0.62$ ~~~ \\
    & enc2-dec2 & $34.38$ & $24.61$ & $32.99$ & $34.38$ & $39.45$ & $30.56$ & $0.00$ & $3.56$ & $-0.58$ ~~~ \\
    \hline
  \end{tabular}
\end{table*}

\subsection{Voice conversion result evaluation}

This part evaluates voice conversion results from FHVAE-APC. The voice conversion process has two relevant speakers: source speaker who offers content information; target speaker is ideally whom the converted utterance sounds like. The voice conversion solution here is the same as in~\cite{10.5555/3294771.3294950}: firstly, the whole end-to-end framework is trained on TIMIT training set; spectrograms from TIMIT core test set are fed into encoders to get $z_1$ and $z_2$; $z_1$ from source speaker and $z_2$ from target speaker are used as input to decoder 1 to get conversion results. For each speaker from TIMIT core test set, one 'SI' utterance is randomly chosen for voice conversion. Thus, 24 utterances from different speakers are chosen in total. Many-to-many voice conversion is processed in pairs between these 24 speakers, from which 576 converted utterances are generated. Generated conversion audio results are assessed on speech recognition and speaker verification aspects. 

\subsubsection{Speech recognition}
A speech recognition system from the Kaldi toolbox is used here \cite{povey2011kaldi}. The acoustic features used are 12-dimensional MFCC. Results are shown in Table~\ref{tab:vc speech-rec}. From Table~\ref{tab:vc speech-rec}, we find that all these four frameworks show similar performance.
'FHVAE-APC enc2-dec2' slightly outperforms other frameworks. The second best framework is baseline, and then is 'FHVAE-APC enc1-dec1' and 'FHVAE-APC enc2-dec1'.

\subsubsection{Speaker verification}
The speaker verification system used in this experiment is supervised Gaussian Mixture Model (GMM), which is initialized according to posteriors from the speech recognition system~\cite{snyder2015time}. Based on this supervised GMM, an i-vector extractor is made, and probabilistic linear discriminant analysis (PLDA) is used for scoring according to i-vector. 12-d MFCC is used as input, while EER as metric for evaluation. 

To assess the voice conversion performance, the experiment has been split into two parts. Experiment A evaluates how much converted utterances sound like the source speaker. Specifically, converted utterances are compared to all speakers except the target speaker in experiment A, and the source speaker of this converted utterance is deemed the true speaker to calculate EER. Experiment B calculates the similarity between converted utterances and target speaker. Details are nearly same as in experiment A, but converted utterance compared with all speakers except source speaker, and target speaker of converted utterance is used as true speaker in EER calculation. $EER_A$ and $EER_B$ are used to denote results from experiment A and B separately. 
To give a total impression of different framework performance, we use a normalized difference as eq.~\eqref{eq: eer_r} shows~\cite{alma9920939670705762}: 
\begin{equation}
    ND = \frac{EER_B-EER_A}{EER_C}
\label{eq: eer_r}
\end{equation}

Since $EER_A$ and $EER_B$ show how similarly the converted utterances sound to the source speaker and target speaker, respectively, the smaller numerator of eq.~\eqref{eq: eer_r} RHS signifies better conversion performance. And to avoid the deviation from speaker verification system, the difference between $EER_A$ and $EER_B$ is normalized by EER result for TIMIT core test set utterances from this speaker verification system. The normalizer is denoted $EER_C$, and $ND$ is used as evaluation metric. Smaller $ND$ shows better performance of this conversion system.

Table \ref{tab:vc speaker-recognition} shows speaker verification results. The EER results for conversion between female and female, between male and male and between all speakers are listed, respectively, with header 'Female', 'Male' and 'All'. $EER_C$ is listed in the row of 'TIMIT core test set'. $ND$, as the overall evaluation metric, are listed in the three rightmost columns. This table shows that: 'FHVAE-APC enc2-dec1' has best performance on same-gender conversion; 'FHVAE' has best performance on all-speaker (i.e. same gender or opposite gender) conversion; the differences between 'FHVAE', 'FHVAE-APC enc2-dec1' and 'FHVAE-APC enc2-dec2' in all cases are small.

\section{Conclusion}
\label{sec:conclusion}
Inspired by the strong feature extraction ability of APC, this paper proposed an FHVAE-APC framework by adding an additional predictive decoder to the FHVAE framework. Three different versions of FHVAE-APC have been explored to evaluate how different structures in encoders and decoders impact performance. Speech recognition and speaker recognition experiments have been utilized to measure performance of latent sequential and segmental features and voice conversion utterances from different frameworks. Results show that the proposed FHVAE-APC has advantages, compared to FHVAE, on feature extraction, both on segmental features and sequential features. For voice conversion, similar performance is observed. 


\bibliographystyle{IEEEbib}
\bibliography{refs}

\end{document}